# Room-temperature van der Waals magnetoresistive memories with data writing by orbital current in the Weyl semimetal TaIrTe$_4$

Dong Li [1,2,*], Xing-Yu Liu,[1,*] Zhen-Cun Pan,[1] An-Qi Wang,[1] Jiantian Zhang,[3] Peng Yu,[3] and Zhi-Min Liao [1,4,†]

[1]*State Key Laboratory for Mesoscopic Physics and Frontiers Science Center for Nano-optoelectronics, School of Physics,*
*Peking University, Beijing 100871, China*
[2]*Academy for Advanced Interdisciplinary Studies, Peking University, Beijing 100871, China*
[3]*State Key Laboratory of Optoelectronic Materials and Technologies, School of Materials Science and Engineering,*
*Sun Yat-sen University, Guangzhou 510275, China*
[4]*Hefei National Laboratory, Hefei 230088, China*

Current-induced out of plane magnetization has been utilized for field-free switching of ferromagnets with perpendicular magnetic anisotropy. Identifying systems capable of energy-efficiently converting charge currents into out of plane orbit- or spin-polarized currents is crucial for advancing magnetic memory technologies. Here we introduce the Berry curvature dipole as a key evaluation factor, directly measurable through nonlinear Hall effects. In the Weyl semimetal TaIrTe$_4$ used in our experiments, applying a current parallel to the Berry curvature dipole results in out of plane orbital magnetization, which governs the field-free perpendicular magnetization switching in TaIrTe$_4$/Fe$_3$GaTe$_2$ heterostructures. Notably, all-electric control of van der Waals magnetoresistive memory at room temperature has been achieved with a low critical current density $\sim 2 \times 10^6$ A/cm$^2$ for data writing. Our findings reveal the connection between nonlinear Hall effects and field-free magnetization switching, highlighting the potential of the Berry curvature dipole in advancing orbitronics.

## I. INTRODUCTION

In the rapidly evolving landscape of artificial intelligence, there is a pressing demand for memory technologies that exhibit nonvolatility, low energy consumption, and high-density integration [1–3]. Magnetoresistive random-access memory (MRAM) stands out as a promising candidate due to its swift access time and robust endurance [4,5]. To further scale down MRAM, the development of magnetic tunnel junctions (MTJs) with perpendicular magnetic anisotropy (PMA) is crucial [6]. However, achieving deterministic field-free switching of perpendicular magnetization (PM) remains a significant challenge, often requiring external magnetic fields or interlayer engineering in conventional heavy metal/ferromagnet bilayers using spin-orbit torque (SOT) mechanisms [7–11]. Transition metal dichalcogenides (TMDs) in the orthorhombic ($T_d$) phase, such as $T_d$-WTe$_2$ and $T_d$-TaIrTe$_4$, exhibit broken twofold rotation symmetry on the surface, enabling the conversion of charge currents to $z$-orbit/spin-polarized currents [12–18]. The field-free switching of PM has been demonstrated utilizing $T_d$-WTe$_2$ and $T_d$-TaIrTe$_4$, attributed to the generation of out of plane antidamping-like torque [14–18]. When considering electrons confined in the two-dimensional (2D) plane, the orbital angular momentum and orbital magnetic moment exclusively possess $z$ components [19–22]. The orbital Edelstein effect [23,24], corresponding to current-induced orbital magnetization, can make a substantial contribution to the antidamping-like torque in the adjacent ferromagnetic layer [15,18,25–27]. In particular, recent revelations regarding the connection between orbital magnetic moment and Berry curvature [28], coupled with the discovery of current-induced orbital magnetization in systems with a Berry curvature dipole (BCD) [29–31], provide an alternative origin of the out of plane antidamping-like torque. The nonzero BCD in both WTe$_2$ and TaIrTe$_4$ has been revealed by measurements of the nonlinear Hall effect (NLHE) [32–34]. Therefore, it is highly desirable to investigate the relationship between field-free PM switching and NLHE to elucidate the underlying mechanisms. However, due to the small BCD and low nonlinear susceptibility, the NLHE can only be observed at low temperatures (<100 K) in WTe$_2$ [32], while magnetization switching generally requires a relatively high temperature. This makes the explicit relationship between NLHE and field-free PM switching remain elusive.

Here, we focus on few-layer $T_d$-TaIrTe$_4$, possessing a larger Berry curvature dipole than $T_d$-WTe$_2$ and exhibiting a pronounced NLHE at room temperature, which enables field-free deterministic PM switching in TaIrTe$_4$/Fe$_3$GaTe$_2$ van der Waals (vdW) heterostructures. The current-induced orbital magnetization in $T_d$-TaIrTe$_4$ is calculated to be a dominant contribution to the generation of the out of plane antidamping-like torque in Fe$_3$GaTe$_2$. By integrating a TaIrTe$_4$ thin layer with the Fe$_3$GaTe$_2$/h-BN/Fe$_3$GaTe$_2$ MTJ, the magnetoresistive memory achieves field-free data writing and reading at room temperature, promoting the advancement of energy-efficient memory technologies.

[*]These authors contributed equally to this work.
[†]Contact author: liaozm@pku.edu.cn

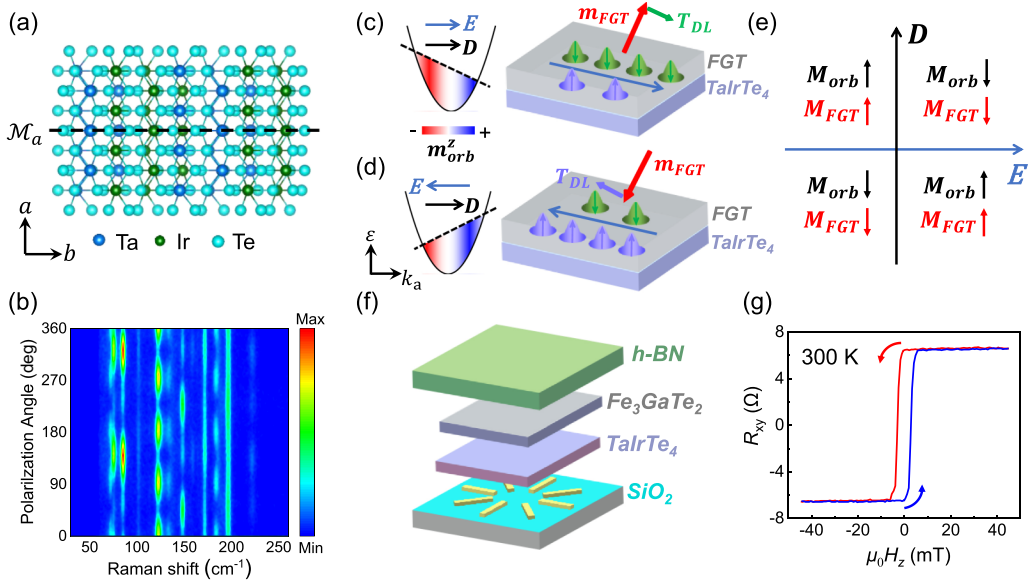

FIG. 1. (a) Crystal structure of few-layer TaIrTe$_4$ viewed from the top. The dashed line represents the mirror plane $\mathcal{M}_a$. (b) Angle-resolved polarized Raman spectral intensity of a thin TaIrTe$_4$ flake. (c), (d) Left column: Illustration of the current-induced polarization of orbital magnetic moment in TaIrTe$_4$ with Berry curvature dipole. Right column: In TaIrTe$_4$/FGT heterostructure, the current-induced out of plane antidamping-like torque drives magnetization switching of FGT. (e) Schematic depiction of orbital magnetization ($M_{orb}$) and FGT magnetization ($M_{FGT}$) within the $D-E$ coordinate diagram. (f) Schematic illustration of a TaIrTe$_4$/Fe$_3$GaTe$_2$ vdW heterostructure device. (g) Hall resistance of TaIrTe$_4$ (4.5 nm)/Fe$_3$GaTe$_2$ (7.2 nm) heterostructure in device B at 300 K.

## II. METHODS

*Device fabrication.* TaIrTe$_4$, h-BN, and Fe$_3$GaTe$_2$ flakes were prepared by mechanical exfoliation onto SiO$_2$/Si substrates, identified by optical contrast. We patterned Ti/Au electrodes (∼12 nm thick) onto an individual SiO$_2$/Si substrate through *e*-beam lithography, metal deposition, and lift-off. The van der Waals heterostructures were made using dry transfer techniques. The capping h-BN, Fe$_3$GaTe$_2$, and TaIrTe$_4$ layers were successively picked up and deposited onto the Ti/Au electrodes to fabricate TaIrTe$_4$/Fe$_3$GaTe$_2$ heterostructure devices. For the magnetoresistive memory device, the flakes were picked up in the following order before being transferred onto the Ti/Au electrodes: capping h-BN, top Fe$_3$GaTe$_2$, thin h-BN, bottom Fe$_3$GaTe$_2$, and TaIrTe$_4$. The whole exfoliation and transfer processes were done in an argon-filled glove box with O$_2$ and H$_2$O content below 0.01 parts per million to avoid sample degradation.

*Transport measurements.* The devices were measured in an Oxford cryostat with a variable temperature insert and a superconducting magnet. Stanford Research Systems SR830 and SR865A lock-in amplifiers were used to measure the first- and second-harmonic voltage signals with a frequency of $\omega = 17.777$ Hz, unless otherwise specified. A Keithley 6221 current source and a Keithley 2182A nanovoltmeter were used for current pulse induced magnetization switching and anomalous Hall effect (AHE) loop shift experiments. For current pulse induced magnetization switching measurements, the used current pulse $I_p$ was a square-wave current pulse with varying magnitude and a width of 60 μs. After each $I_p$ was applied and then removed, the Hall resistance ($R_{xy}$) or the resistance of the MTJ ($R_{MTJ}$) was measured. AHE loop shift measurements were performed by measuring $R_{xy}(H_z)$, where at each perpendicular field $H_z$, a pulse current $I_p$ (60 μs long) was applied and then $R_{xy}$ was measured by applying a 1 μA bias ac current.

*First-principles calculations.* We performed *ab initio* density-functional theory (DFT) calculations on a five-layer $T_d$-TaIrTe$_4$ slab and then constructed a DFT-based tight-binding model Hamilton, where the tight-binding model matrix elements were calculated by projecting onto the Wannier orbitals [35,36]. The *d* orbitals of Ta atoms, *d* orbitals of Ir atoms, and *p* orbitals of Te atoms were used to construct Wannier functions using the WANNIER90 code [37], without performing the procedure for maximizing localization. The Berry curvature dipole and orbital (spin) magnetoelectric coefficients were calculated using the WANNIERBERRI code [38], where the integral over the two-dimensional Brillouin zone was taken on the 1501 × 501 fine grid.

## III. RESULTS AND DISCUSSION

The crystal structure of $T_d$-TaIrTe$_4$ belongs to the *Pmn*2$_1$ space group [39]. On the surface, there is a distinct mirror symmetry ($\mathcal{M}_a$) with a single mirror line along the crystal *b* axis, as shown in Fig. 1(a). Figure 1(b) plots the angular dependence of polarized Raman intensity spectra measured with 532 nm excitation wavelengths through a linearly polarized solid-state laser beam. The integrated intensities of a Raman peak at around 146 cm$^{-1}$ are used to identify the *a* axis of TaIrTe$_4$ [40]. The existence of a notable Berry curvature dipole (*D*) along the *a* axis at room temperature is identified by the NLHE measurements (see Appendix A). Application of current along the *a* axis can induce polarization of the orbital magnetic moment ($m_{orb}$), resulting in the orbital magnetization ($M_{orb}$), as illustrated in the left column of Figs. 1(c) and 1(d). When TaIrTe$_4$ is in proximity to a ferromagnetic layer with PMA, such as Fe$_3$GaTe$_2$ (FGT) [41] used in this work, the induced orbital magnetization generates an out of plane antidamping-like torque ($\boldsymbol{T}_{\text{DL}}$) in FGT, given by $\boldsymbol{m}_{\text{FGT}} \times (\boldsymbol{m}_{\text{orb}} \times \boldsymbol{m}_{\text{FGT}})$ [42], where $\boldsymbol{m}_{\text{FGT}}$ is the magnetic moment of the FGT layer. The $\boldsymbol{T}_{\text{DL}}$ enforces the magnetization of FGT ($\boldsymbol{M}_{\text{FGT}}$) to align preferentially with the direction of $\boldsymbol{M}_{\text{orb}}$. Consequently, field-free PM switching is achievable. Moreover, upon reversing the polarity of the applied current, the directions of $\boldsymbol{M}_{\text{orb}}$ and $\boldsymbol{T}_{\text{DL}}$ are also reversed, as illustrated in the right column of Figs. 1(c) and 1(d). Therefore, the terminal states of $\boldsymbol{M}_{\text{FGT}}$ are determined by $\boldsymbol{M}_{\text{orb}}$, which is controlled by the direction of $-(\boldsymbol{D} \cdot \boldsymbol{E})\hat{z}$, as outlined in Fig. 1(e). The TaIrTe$_4$/FGT heterostructure is illustrated in Fig. 1(f). The alignment of the crystalline axes of TaIrTe$_4$ with the electrodes is ensured by identifying long, straight edges and confirmed through polarized Raman spectroscopy. We conduct measurements on five TaIrTe$_4$/FGT devices (devices B–F). The FGT shows a PMA characteristic at room temperature [Fig. 1(g)].

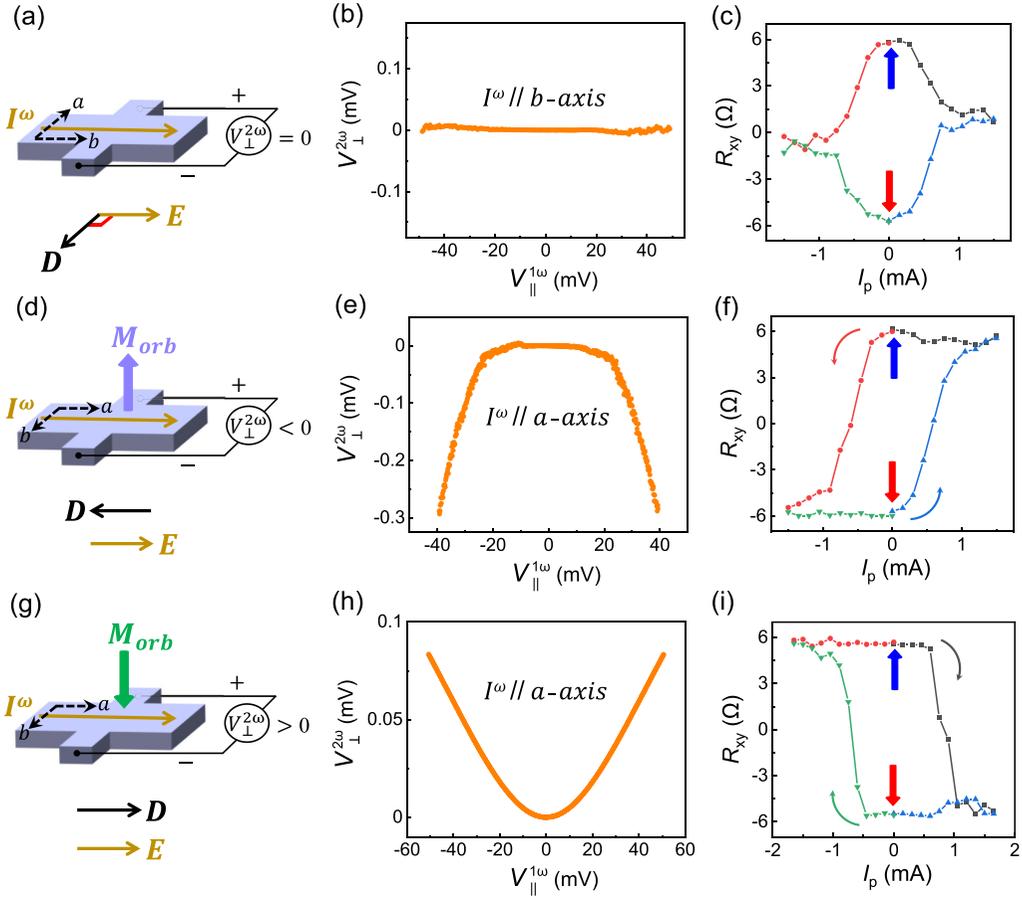

FIG. 2. (a), (d), (g) Depiction of the interplay between the Berry curvature dipole $\boldsymbol{D}$, current-induced orbital magnetization $\boldsymbol{M}_{\text{orb}}$ and second-harmonic Hall voltage $V_{\perp}^{2\omega}$. (b), (e), (h) The NLHE measured in TaIrTe$_4$ at 300 K with $I^{\omega}$ along (b) the $b$ axis in device B, (e) the $a$ axis in device B, and (h) the $a$ axis in device D, respectively. $V_{\parallel}^{1\omega} = I^{\omega} R_{\parallel}$, where $R_{\parallel}$ is the longitudinal resistance. (c), (f), (i) Magnetization switching, reflected by the Hall resistance measurements, induced by the current pulse $I_p$ applied in TaIrTe$_4$ along (c) the $b$ axis in device B, (f) the $a$ axis in device B, and (i) the $a$ axis in device D, respectively.

The NLHE is proportional to $(\boldsymbol{D} \cdot \boldsymbol{E})\hat{z} \times \boldsymbol{E}$ [43] and is also related to $-\boldsymbol{M}_{\text{orb}} \times \boldsymbol{E}$, providing a measurable connection between the Berry curvature dipole and the magnetization switching. Given that electrons are the dominant carriers in our devices (see Appendix B), the direction of $\boldsymbol{M}_{\text{orb}}$ can be obtained through the second-order Hall voltage ($V_{\perp}^{2\omega}$), as illustrated in Figs. 2(a), 2(d), and 2(g). In Fig. 2(b), upon applying an ac current ($I^{\omega}$) along the $b$ axis of TaIrTe$_4$ in device B, $V_{\perp}^{2\omega}$ is nearly zero, indicating that $\boldsymbol{E}$ is perpendicular to $\boldsymbol{D}$ [Fig. 2(a)]. In this case, the electric field does not induce an out of plane orbital magnetization and the out of plane antidamping-like torque. When a dc current pulse ($I_p$) is applied, no deterministic switching is observed, and demagnetized states ($R_{xy} \sim 0$) can be achieved [Fig. 2(c)]. This behavior stems from the conventional SOT generated by in-plane spin accumulations, which propels the magnetic moments of FGT toward the device plane. Upon removal of $I_p$, each magnetic domain adopts a random orientation, either upward or downward, leading to the demagnetized states of FGT.

In Fig. 2(e), $V_{\perp}^{2\omega} < 0$ is observed when applying $I^{\omega}$ along the $a$ axis. Notably, the polarity of $V_{\perp}^{2\omega}$ remains unchanged when exchanging the source-drain electrodes of the bias current, consistent with $V_{\perp}^{2\omega}$ having a quadratic dependence on $I^{\omega}$. As depicted in Fig. 2(d), when applying a positive $\boldsymbol{E}$, a measured $V_{\perp}^{2\omega} < 0$ corresponds to $\boldsymbol{D}$ being antiparallel to $\boldsymbol{E}$ and $\boldsymbol{M}_{\text{orb}}$ aligning along the $+\hat{z}$ direction. As shown in

TABLE I. Summary of temperature ($T$), external magnetic field ($H_{ext}$), critical current density ($J_c$), and antidamping-like torque efficiency ($\xi_{DL}$) in current-induced perpendicular magnetization switching for different material systems.

| Systems | $T$ | $H_{ext}$ | $J_c$ (A cm$^{-2}$) | $\xi_{DL}$ | Reference |
|---|---|---|---|---|---|
| Pt/Fe$_3$GeTe$_2$ | 180 K | 3000 Oe | $2.5 \times 10^7$ | 0.14 | Alghamdi et al. [44] |
| Pt/Fe$_3$GeTe$_2$ | 100 K | 500 Oe | $9.3 \times 10^6$ | 0.12 | Wang et al. [45] |
| (Bi, Sb)$_2$Te$_3$/Fe$_3$GeTe$_2$ | 180 K | 1000 Oe | $1.7 \times 10^6$ | — | Fujimura et al. [46] |
| WTe$_2$/Fe$_3$GeTe$_2$ | 170 K | Field-free | $7.9 \times 10^6$ | — | Kao et al. [14] |
| WTe$_2$/Fe$_3$GeTe$_2$ | 120 K | Field-free | $6.5 \times 10^6$ | — | Ye et al. [15] |
| WTe$_2$/Fe$_3$GaTe$_2$ | 300 K | Field-free | $3 \times 10^6$ | — | Pan et al. [18] |
| Pt/Fe$_3$GaTe$_2$ | 300 K | 350 Oe | $4.8 \times 10^6$ | — | Yun et al. [47] |
| Bi$_{0.9}$Sb$_{0.1}$/MnGa | 300 K | 3500 Oe | $1.5 \times 10^5$ | 52 | Khang et al. [48] |
| Bi$_2$Se$_3$/Gd$_x$(FeCo)$_{1-x}$ | 300 K | 100 Oe | $1.2 \times 10^6$ | 0.13 | Wu et al. [49] |
| TaIrTe$_4$/CoFeB | 300 K | Field-free | $7.6 \times 10^6$ | 0.048 | Liu et al. [16] |
| TaIrTe$_4$/Ti/CoFeB | 300 K | Field-free | $2.4 \times 10^6$ | 0.05 | Zhang et al. [17] |
| TaIrTe$_4$/Fe$_3$GaTe$_2$ | 300 K | Field-free | $1.3 \times 10^6$ | 0.16 | This work |

Fig. 2(f), applying $I_p$ achieves a fully field-free deterministic PM switching at 300 K. Additionally, employing a series of positive and negative current pulses with an amplitude of 1.5 mA can reliably switch the magnetization of FGT (see Appendix C). The critical current density is $2.36 \times 10^6$ A/cm$^2$ ($1.38 \times 10^6$ A/cm$^2$ for device D), which compares favorably with conventional heavy metal/ferromagnetic systems and WTe$_2$/Fe$_3$GeTe$_2$ heterostructures (see Table I for detailed comparison).

Differently, with the measurement frame kept unchanged, a positive $V_\perp^{2\omega}$ is observed when applying $I^\omega$ along the $a$ axis of TaIrTe$_4$ in device D [Fig. 2(h)]. As shown in Fig. 2(g), when applying a positive $E$, a measured $V_\perp^{2\omega} > 0$ corresponds to $D$ being parallel to $E$ and the induced $M_{orb}$ aligning along the $-\hat{z}$ direction, resulting in an opposite polarity of the PM switching [Fig. 2(i)]. Taken collectively, the polarity of the $R_{xy} - I_p$ loops (clockwise or anticlockwise) can be determined simply by the sign of the nonlinear Hall voltage under the same measurement configuration. The consistent results are also reproducible in other devices (see Appendix D).

Besides measurements along the $a$ and $b$ axes, we conduct additional experiments with currents applied along the directions of $\theta = 45°$ and $135°$. The angle $\theta$ is defined as the angle between $I_p$ and the $a$ axis, as depicted in Fig. 3(a). Due to the nonzero current components along the $a$ axis, both the NLHE and field-free PM switching are observed at $\theta = 45°$ and $135°$. At $\theta = 45°$, where the applied $E$ has a component antiparallel to $D$, negative $V_\perp^{2\omega}$ and anticlockwise polarity of the $R_{xy} - I_p$ loop are observed [Figs. 3(b) and 3(c)]. In contrast, at $\theta = 135°$, positive $V_\perp^{2\omega}$ and clockwise polarity in the $R_{xy} - I_p$ loops are observed [Figs. 3(b) and 3(d)], which is consistent with the fact that $E$ has a component parallel to $D$. Furthermore, compared to $\theta = 0°$ [Figs. 2(e) and 2(f)], both the NLHE and the height of the $R_{xy} - I_p$ loops at $\theta = 45°$ and $135°$ are relatively smaller. It indicates that the terminal magnetization of FGT relies on the direction and strength of $-(D \cdot E)\hat{z}$, underscoring the crucial role of current-induced orbital magnetization in the field-free PM switching in TaIrTe$_4$/FGT heterostructures.

To better comprehend the connection between NLHE and field-free PM switching, we perform theoretical analysis and calculations to explore the underlying mechanisms. The electric field induced orbital magnetization can be described as $M_{orb} = (\frac{e^2\tau}{2\hbar^2})\alpha^{orb}E$, where $e$ is the electron charge, $\tau$ is the relaxation time, and $\hbar$ is the reduced Planck's constant. The orbital magnetoelectric coefficient $\alpha^{orb}$ is defined as [23,50]

$$\alpha_{ij}^{orb} = \int_{BZ} \frac{d^2k}{(2\pi)^2} \sum_n m_{nk}^{j,orb} \partial_{k_i}\epsilon_{nk} \partial_{\epsilon_{nk}} f_{nk}^{(0)},$$

where $i$ and $j$ represent $x$, $y$, or $z$; $f_{nk}^{(0)}$ is the Fermi distribution function; $\epsilon_{nk}$ is the energy eigenvalue of the band $n$ at $\mathbf{k}$; and $\mathbf{m}_{nk}^{orb} = \text{Im}\langle \partial_k u_{nk} | \times (H_k - \epsilon_{nk}) | \partial_k u_{nk} \rangle$ is the orbital magnetic moment of each Bloch state. It can be divided into two contributions [51]: $\alpha_{ij}^{orb} = -\mu D_{ij} + \beta_{ij}$. The first term is linear with the chemical potential $\mu$ and proportional to the BCD

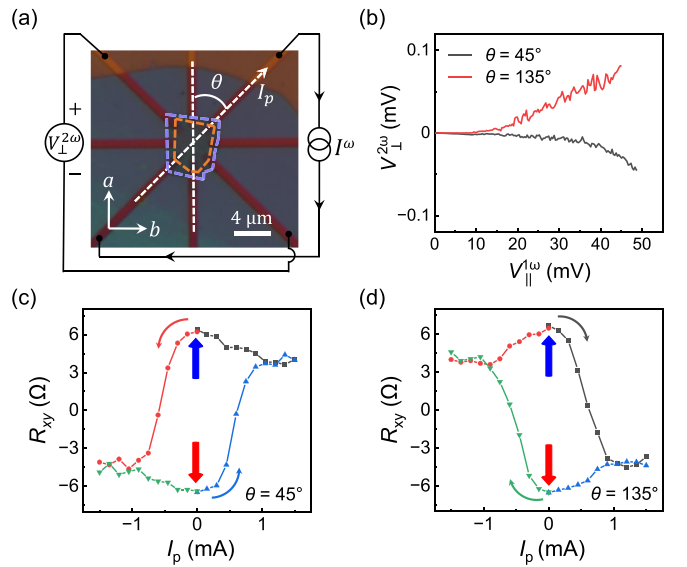

FIG. 3. (a) The optical image of device B, where an angle $\theta$ is defined. (b) The NLHE measured at 300 K with $I^\omega$ along $\theta = 45°$ and $135°$, respectively. (c), (d) Magnetization switching with $I_p$ along (c) $\theta = 45°$ and (d) $\theta = 135°$ at 300 K, respectively.

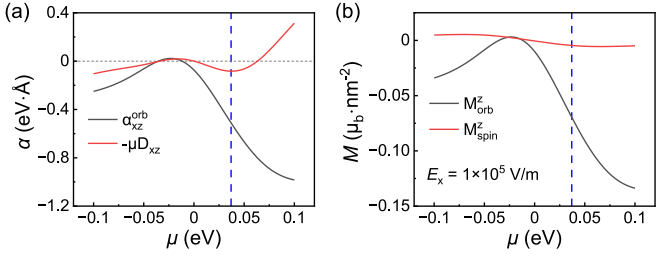

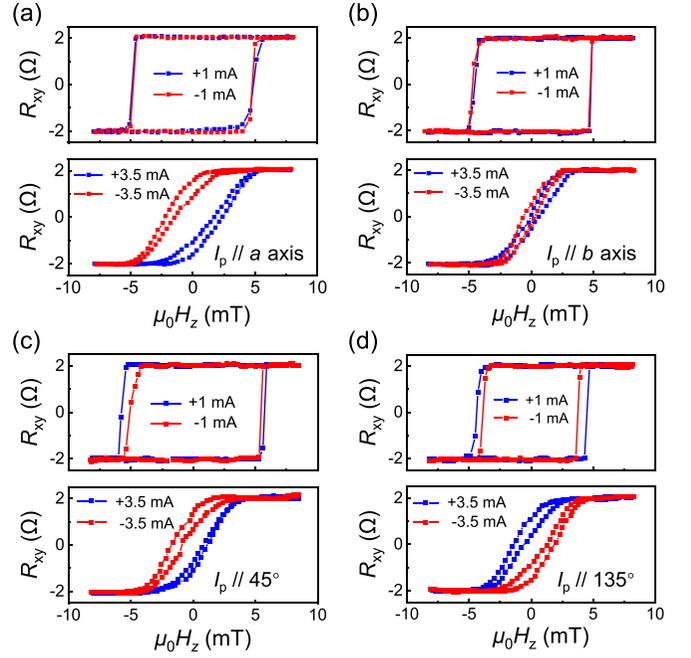

FIG. 4. (a) Calculated orbital magnetoelectric susceptibility $\alpha_{xz}^{\text{orb}}$ and its contribution from BCD ($-\mu D_{xz}$) as a function of chemical potential $\mu$. The blue dashed line indicates the Fermi level ($\mu = 0.037$ eV) of TaIrTe$_4$ in our devices. The temperature used for the calculations is 300 K, which is included in the Fermi-Dirac distribution. The coordinates $x$, $z$ correspond to the $a$ and $c$ axes of TaIrTe$_4$, respectively. (b) Comparison of out of plane orbital magnetization and spin magnetization as a function of $\mu$, with an applied electric field of $10^5$ V/m along the $a$ axis of TaIrTe$_4$. The temperature and relaxation time used for the calculations are 300 K and 1 ps, respectively.

[43],

$$D_{ij} = -\int_{\text{BZ}} \frac{d^2 k}{(2\pi)^2} \sum_n \partial_{k_i} \epsilon_{n\mathbf{k}} \Omega_{n\mathbf{k}}^j \partial_{\epsilon_{n\mathbf{k}}} f_{n\mathbf{k}}^{(0)},$$

where the Berry curvature is formulated as $\mathbf{\Omega}_{n\mathbf{k}} = -\text{Im}\langle \nabla_{\mathbf{k}} u_{n\mathbf{k}}| \times |\nabla_{\mathbf{k}} u_{n\mathbf{k}}\rangle$. The $\beta_{ij}$ item is nonlinear in $\mu$ and emerges from an effective magnetic field possessing symmetries akin to Berry curvature and orbital magnetic moment. For gapped Dirac systems (e.g., strained MoS$_2$ monolayer [29]), the $\alpha_{ij}^{\text{orb}} \approx -\mu D_{ij}$ as long as the chemical potential $\mu$ remains in the vicinity of the Dirac nodes.

In the more complex case of few-layer TaIrTe$_4$, the relationship between Berry curvature dipole and current-induced orbital magnetization is further investigated by first-principles calculations (for more details see Appendix E). Figure 4(a) presents the calculated $\alpha_{xz}^{\text{orb}}$ and its contribution from $-\mu D_{xz}$ as a function of $\mu$ for a thin-layer $T_d$-TaIrTe$_4$ model. Based on the measured carrier densities, the Fermi level of TaIrTe$_4$ is estimated to near $\mu = 0.037$ eV, as indicated by the blue dashed line in Fig. 4(a). Notably, Fig. 4(a) illustrates that $-\mu D_{xz}$ makes a positive contribution to $\alpha_{xz}^{\text{orb}}$ at the chemical potential corresponding to the sample's Fermi level. Furthermore, the second-order NLHE also has a quantum origin linked to the Berry curvature dipole. This suggests that current-induced orbital magnetization and second-order NLHE are interrelated phenomena in thin-layer TaIrTe$_4$.

Nevertheless, the presence of spin-orbit coupling and low crystal symmetry in TaIrTe$_4$ suggests that the current-induced $z$-direction spin polarization also exhibits. To assess the relative contributions of spin and orbital moments to the total out of plane magnetization, we calculate the spin magnetoelectric coefficient, denoted as $\alpha_{xz}^{\text{spin}}$ (see Appendix E). In Fig. 4(b), we plot the out of plane spin and orbital magnetizations as functions of the chemical potential $\mu$, with an external electric field $E_x$ of $10^5$ V/m applied along the $a$ axis of TaIrTe$_4$, respectively. Our calculations reveal that the orbital magnetization surpasses the spin magnetization by a factor of 15 at the Fermi level ($\mu = 0.037$ eV). This underscores the dominant role of orbital magnetization in contributing to current-induced out of plane magnetization, reinforcing its significance in generation of the out of plane antidamping-like torque and field-free PM switching. Therefore, it enables us to establish a connection between NLHE and field-free PM switching in TaIrTe$_4$/Fe$_3$GaTe$_2$ vdW heterostructure devices.

To quantify the strength of the generated out of plane antidamping-like torque, we conduct anomalous Hall effect (AHE) hysteresis loop shift measurements in device F. An antidamping-like torque typically induces a shift in the AHE hysteresis loop to counteract intrinsic damping once the current surpasses a critical threshold [6,52]. As illustrated in the top panel of Fig. 5(a), when applying positive (+1 mA) and negative (−1 mA) pulse currents along the $a$ axis ($\theta = 0°$), the AHE hysteresis loops appear identical for different current polarities, displaying negligible AHE loop shift. However, when $I_p$ increases to +(−) 3.5 mA, the AHE hysteresis loop center shifts to the right (left), as depicted in the bottom panel of Fig. 5(a). In contrast, no significant loop shift is observed for both low (±1 mA) and high (±3.5 mA) current pulse magnitudes when $I_p$ is applied along the $b$ axis ($\theta = 90°$), as shown in Fig. 5(b). It is worth noting that the coercive field of FGT also decreases at high current density, primarily due to Joule heating (see discussions in Appendix F). For $\theta = 135°$, the observed loop shifts are opposite to those at 45° and 0°, showing a left (right) loop shift for a +3.5 (−3.5) mA $I_p$ (see Fig. 5).

The loop shift field $H_{\text{shift}}^+$ ($H_{\text{shift}}^-$) is measured at a positive (negative) pulse current, which is defined as $[H_c^+ + H_c^-]/2$, where $H_c^+$ ($H_c^-$) represents the magnetic field at which the magnetization switches from down to up (up to down). The $H_{\text{shift}}^\pm$ is closely related to the current pulses along different directions with various $\theta$ [Fig. 6(a)]. For $\theta = 90°$ ($b$ axis), the $H_{\text{shift}}^{\pm}$ remains nearly zero for the specified current values. However, for $\theta = 0°$, $45°$, and $135°$, a distinct separation between $H_{\text{shift}}^{+}$ and $H_{\text{shift}}^{-}$ becomes apparent when the current values surpass 1 mA, reaching saturation as the current approaches the critical value (3.5 mA). The average loop shift field ($H_{\text{shift}}$) is estimated by $H_{\text{shift}} = \frac{H_{\text{shift}}^{+} - H_{\text{shift}}^{-}}{2}$. By using $H_{\text{shift}}^{+}$ and $H_{\text{shift}}^{-}$ measured at $\pm 3.5$ mA, the $H_{\text{shift}}$ are obtained for different $\theta$ values, as presented in Fig. 6(b). The data fit well with a cosine function, exhibiting an amplitude of 2 mT for $\theta = 0°$ (a axis). Additionally, we determine the out of plane antidamping-like torque efficiency ($\xi_{\text{DL}}^{z}$) using the formula $H_{\text{shift}}/J_{\text{in}} = \frac{\pi}{2} \frac{\hbar \xi_{\text{DL}}^{z}}{2e\mu_0 M_s t_{\text{FM}}}$ [53,54], where $J_{\text{in}}$ is the input charge current density, $\hbar$ is the reduced Planck's constant, $e$ is the electron charge, $\mu_0$ is the vacuum permeability, $M_s$ is the saturation magnetization of FGT, and $t_{\text{FM}}$ is the thickness of the FGT layer. Utilizing $H_{\text{shift}}/J_{\text{in}} = 2\text{mT}/(4.9 \times 10^6 \text{A/cm}^2)$ obtained from the AHE hysteresis loop shift measurements, $t_{\text{FM}} = 9.4$ nm, and the reported room temperature $M_s = 40.11$ emu/g [41], we obtain a room temperature out of plane torque efficiency of $\xi_{\text{DL}}^{z} \sim 0.16$ for the current along the $a$ axis, surpassing observations in WTe$_2$ [18]. These results highlight TaIrTe$_4$ as a promising candidate for spintronic devices.

FIG. 5. AHE hysteresis loops measured at 290 K in device F with current pulses applied along (a) the $a$ axis ($\theta = 0°$), (b) the $b$ axis ($\theta = 90°$), (c) $\theta = 45°$, and (d) $\theta = 135°$, respectively.

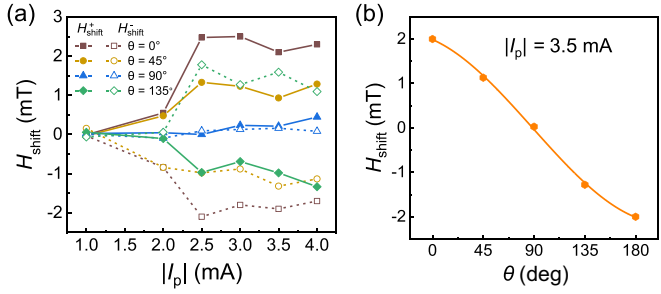

FIG. 6. (a) Angular dependence of the loop shift field ($H_{\text{shift}}^{\pm}$) measured as a function of the pulse current amplitude $|I_p|$ for positive ($H_{\text{shift}}^{+}$, solid line) and negative ($H_{\text{shift}}^{-}$, dashed line) pulse current at 290 K in device F. (b) The hysteresis loop shift field ($H_{\text{shift}}$) measured as a function of $\theta$ for the pulse current magnitude $|I_p| = 3.5$ mA. The solid line represents a cosine fit to the data.

Leveraging the high out of plane torque efficiency in TaIrTe$_4$/FGT heterostructures, we extend our exploration by showcasing a van der Waals magnetoresistive memory. This is achieved through the integration of a TaIrTe$_4$ layer with a magnetic tunnel junction (MTJ) based on FGT/thin h-BN layer/FGT, as depicted in Fig. 7(a). The optical image of the device is shown in the left column of Fig. 7(b), featuring stacked structures from top to bottom: capping h-BN (25 nm thick), top FGT layer (15 nm), thin h-BN tunneling layer (2.5 nm), bottom FGT layer (5.8 nm), and TaIrTe$_4$ layer (8.2 nm). The polar diagram in the right column of Fig. 7(b) illustrates the angle dependence of the intensity of the Raman peak at 146 cm$^{-1}$ obtained from polarized Raman spectrum measurements, confirming that the applied writing current direction (indicated by the arrow in the optical image) aligns with the $a$ axis of TaIrTe$_4$.

Applying a bias of 100 nA across the MTJ, the tunneling resistance ($R_{\text{MTJ}}$) exhibits variations between high- ($R_{\text{AP}}$) and low- ($R_{\text{P}}$) resistance states as the sweeping magnetic field induces antiparallel and parallel magnetization alignments of the two FGT layers, as shown in Fig. 7(c). Note that the shoulders of the tunneling resistance in Fig. 7(c) are caused by multiple magnetic domains in one of the FGT layers, resulting from nonuniformity in its thickness or degeneration during the complicated device fabrication process. The tunneling magnetoresistance (TMR) ratio, calculated by $100\% \times \frac{R_{AP}-R_P}{R_P}$, is ~2.6% at 290 K [Fig. 7(c)]. Using the TMR ratio defined as $\frac{2P^2}{1-P^2}$, where $P$ represents the spin polarizability at the Fermi level for free (bottom) and reference (top) $Fe_3GaTe_2$ layers [42], the spin polarizabilities of $Fe_3GaTe_2$ layers at 290 K are obtained to be 11.3%. As presented in Fig. 8(a), the resistance of the MTJ as a function of magnetic field at various temperatures was also measured. Both TMR ratio and spin polarizability are reduced by raising the temperature [Fig. 8(b)].

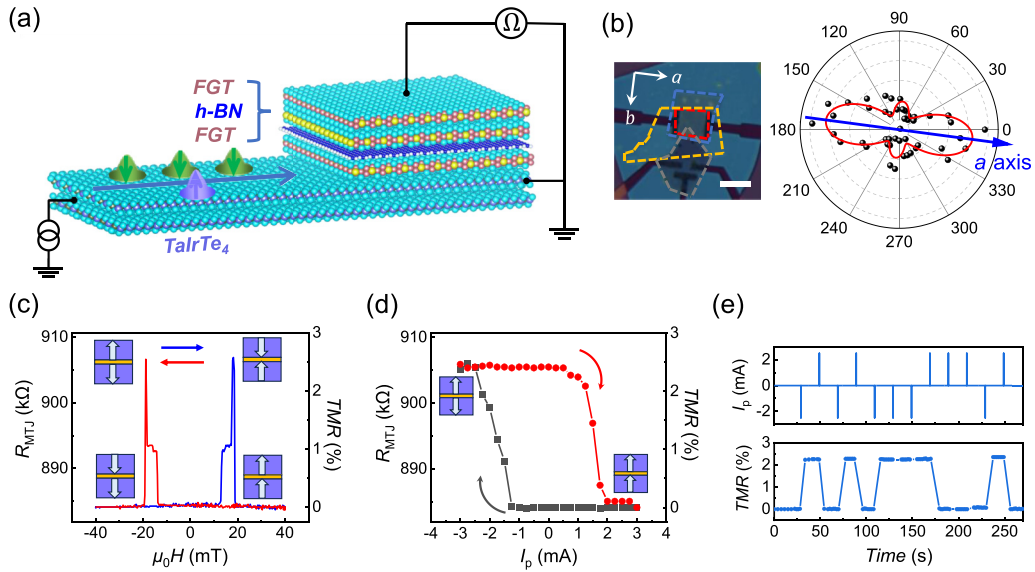

FIG. 7. (a) Schematic of the TaIrTe$_4$/Fe$_3$GaTe$_2$/h-BN/Fe$_3$GaTe$_2$ vdW heterostructure for magnetoresistive memory device. (b) Left: Optical image of the magnetoresistive memory device. The top Fe$_3$GaTe$_2$, thin h-BN, bottom Fe$_3$GaTe$_2$, and TaIrTe$_4$ flakes are distinctly labeled by gray, orange, red, and blue dashed borders, respectively. The scale bar is 6 µm. The electrodes are configured for current pulse application along the $a$ axis of TaIrTe$_4$. Right: Polarized Raman spectroscopy to identify the crystalline $a$ axis of TaIrTe$_4$. (c) Resistance of MTJ and TMR ratio versus the out of plane magnetic field of the device under a 100 nA bias ac current at 290 K. (d) Field-free switching between high- and low-resistance states of the tunneling resistance, achieved by applying writing current pulses at 290 K. (e) Upper: Application of a train of current pulses with an amplitude of ±2.5 mA at 290 K. Lower: Corresponding changes in TMR ratio.

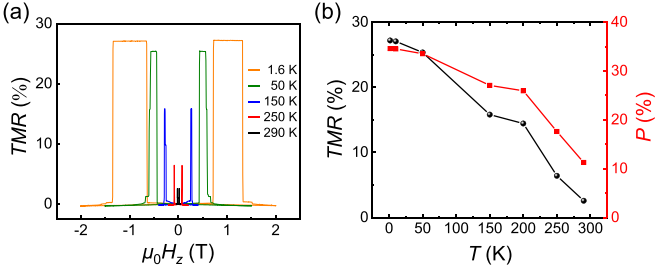

FIG. 8. (a) TMR of the magnetoresistive memory device measured at various temperatures. (b) Temperature dependence of the TMR ratio (black) and spin polarizability (red).

The magnetoresistive memory's data writing is accomplished by field-free switching the magnetization of the bottom FGT layer through the application of current pulses in $TaIrTe_4$. As illustrated in Fig. 7(d), the transition of $R_{MTJ}$ between high- and low-resistance states is achieved by applying current pulses in $TaIrTe_4$. Furthermore, utilizing a series of ±2.5 mA writing current pulses, the state transition between "0" (high-$R_{MTJ}$ state) and "1" (low-$R_{MTJ}$ state) demonstrates robustness and nonvolatility [Fig. 7(e)]. The critical current density ($J_c$) and the power consumption density for data writing is $2.53 \times 10^6$ A/cm² and ~1.35 fJ/nm², respectively. Under the condition of a 2.5 mA current pulse for data writing, thermal effects do not lead to the disappearance of the ferromagnetic properties of FGT (see discussions in Appendix F), indicating that the device temperature is still below the Curie temperature of $Fe_3GaTe_2$. It is anticipated that this device could potentially operate at even higher temperatures beyond our measurement system's limit.

In summary, we have elucidated a profound connection between field-free PM switching and NLHE in $TaIrTe_4/Fe_3GaTe_2$ vdW heterostructures. This correlation stems from the current-induced orbital magnetization in $TaIrTe_4$, governed by the mechanism of the Berry curvature dipole. Our experimental results demonstrate that a simple measurement of the NLHE provides insight into the final magnetization state of the ferromagnetic layer under current drive. This presents a convenient approach for identifying material systems with lower critical current density and higher torque efficiency for PM switching. Building upon this profound understanding, we demonstrate field-free data writing and reading in magnetoresistive memories at room temperature based on the stacking of the $TaIrTe_4$ layer with a $Fe_3GaTe_2$-based vdW MTJ, showcasing characteristics such as low power consumption and nonvolatility. Our work not only advances the understanding of utilizing the Berry curvature dipole for achieving PM switching but also highlights its potential in spintronics for the development of low-power, nonvolatile, and compact devices.

## ACKNOWLEDGMENTS

This work was supported by the National Natural Science Foundation of China (Grants No. 62425401 and No. 62321004) and the Innovation Program for Quantum Science and Technology (Grant No. 2021ZD0302403). P.Y. was supported by the National Natural Science Foundation of China (Grant No. 22175203) and the Natural Science Foundation of Guangdong Province (Grant No. 2022B1515020065).

## APPENDIX A: ANGLE-DEPENDENT LONGITUDINAL RESISTANCE AND NONLINEAR HALL EFFECTS

We fabricated a few-layer $TaIrTe_4$ (with a thickness of 8 nm) device (device A) using circular disk electrodes [inset in Fig. 9(a)], in which we can perform angle-resolved electrical measurements with electrodes aligned with the crystal axes. As shown in Fig. 9(a), the longitudinal resistance exhibits twofold angular dependence with the $a$ axis being the most conductive direction. This is consistent with the crystal symmetry of few-layer TaIrTe$_4$ and can be described as $R_{xx}(\theta) = R_a\cos^2\theta + R_b\sin^2\theta$, where $R_a$ and $R_b$ are the resistances along the crystal $a$ and $b$ axes, respectively. To detect the BCD in TaIrTe$_4$, we also measured the second-order NLHE in this device. We applied a harmonic current $I^\omega$ ($\omega = 17.777$ Hz) and measured both the first-harmonic $\omega$ and second-harmonic $2\omega$ frequencies of the longitudinal and transverse voltages using lock-in amplifiers [inset in Fig. 9(b)] at 300 K. As seen from Fig. 9(b), when applying an $I^\omega$ along the $a$ axis, a clear signal for the second-harmonic transverse voltage $V_\perp^{2\omega}$ can be observed. The $V_\perp^{2\omega}$ scales quadratically with the first-harmonic longitudinal voltage $V_\parallel^{1\omega}$, that is $V_\perp^{2\omega} \propto (V_\parallel^{1\omega})^2$, indicating significant NLHE in TaIrTe$_4$ at room temperature. However, in contrast to the $a$ axis, the second-harmonic voltage $V_\perp^{2\omega}$ is almost zero with a current applied along the $b$ axis. This is attributed to the BCD along the $a$ axis of TaIrTe$_4$.

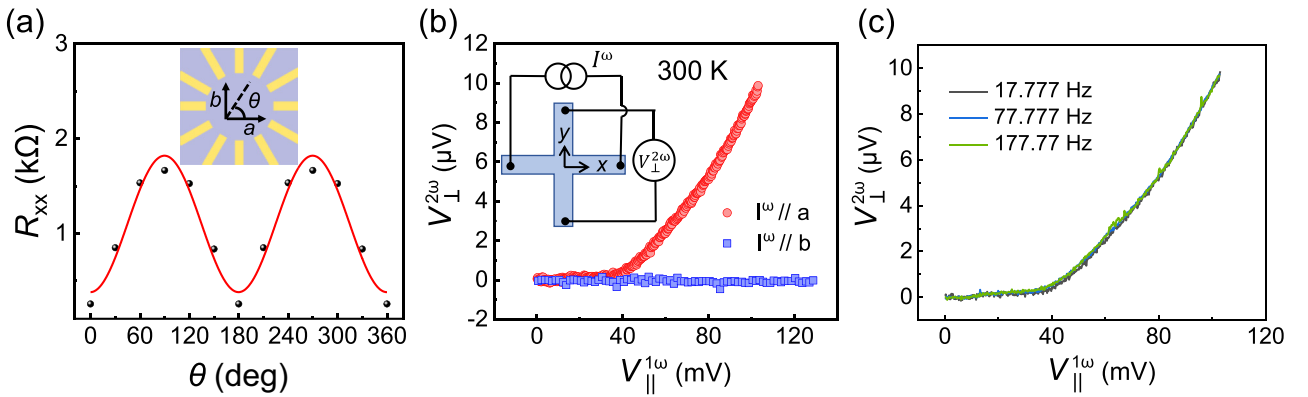

FIG. 9. (a) Resistance anisotropy at 300 K in device A. The $\theta$ is defined as the angle relative to the $a$ axis. (b) The second-harmonic Hall voltage $V_\perp^{2\omega}$ measured as a function of the first-harmonic longitudinal voltage $V_\parallel^{1\omega}$ when applying a harmonic current $I^\omega$ along the $a$ and $b$ axes of $TaIrTe_4$ in device A at 300 K, respectively. (c) Second-harmonic Hall voltage $V_\perp^{2\omega}$ as a function of $V_\parallel^{1\omega}$ for different driving frequencies in device A. The harmonic current $I^\omega$ is applied along the $a$ axis of $TaIrTe_4$.

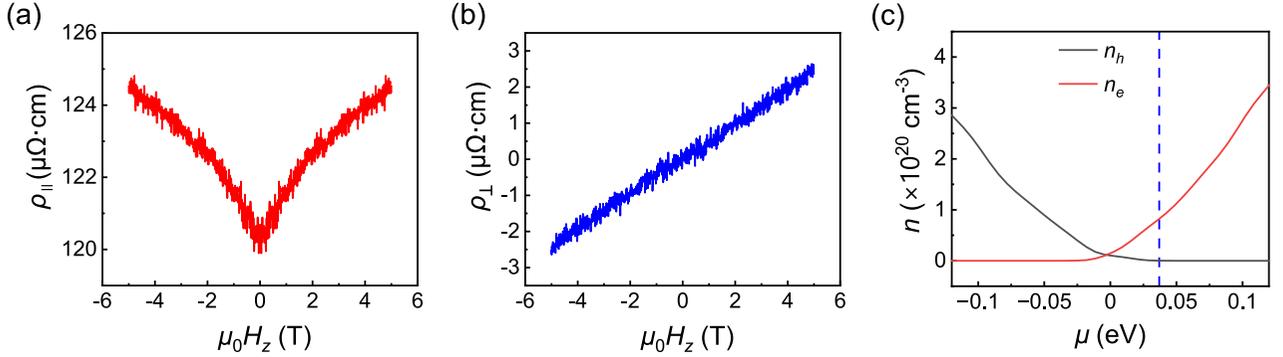

FIG. 10. (a) Magnetic-field dependence of the longitudinal resistivity $\rho_\parallel$. (b) Magnetic-field dependence of the transverse resistivity $\rho_\perp$. (c) The corresponding carrier densities of electron ($n_e$) and hole ($n_h$) in the $T_d$-TaIrTe$_4$ as a function of chemical potential $\mu$. Blue dashed line denotes the position of Fermi level ($\mu = 0.037$ eV) in our TaIrTe$_4$ device.

Figure 9(c) shows the second-harmonic transverse voltage ($V_\perp^{2\omega}$) as a function of the first-harmonic longitudinal voltage $V_\parallel^{1\omega}$ for different driving frequencies in device A. The harmonic current $I^\omega$ is applied along the $a$ axis of TaIrTe$_4$. No frequency dependence is observed in the frequency range we have investigated (17.777–177.77 Hz). This validates the essential property of the BCD and excludes potential measurement artifacts such as spurious capacitive coupling.

## APPENDIX B: EXPERIMENTAL AND CALCULATED CARRIER DENSITIES

We performed Hall measurements under an out of plane magnetic field to extract the density and mobility of the electrons and holes. Figures 10(a) and 10(b) show the magnetic-field dependence of the longitudinal ($\rho_\parallel$) and transverse ($\rho_\perp$) resistivity for device A, respectively. A semiclassical two-carrier model [55] is employed to describe these dependences. We extract the carrier densities, about $12 \times 10^{20}$ cm$^{-3}$ (electron) and about $0.025 \times 10^{20}$ cm$^{-3}$ (hole).

To compare with experimental carrier density values, based on first-principles calculations, the carrier densities $n_e$ and $n_h$ are calculated by [34]

$$n_e = \int_{\epsilon_c}^{\infty} g_e(E) f_0(E - \mu_F) dE,$$
$$n_h = \int_{-\infty}^{\epsilon_v} g_h(E) f_0(\mu_F - E) dE,$$
(B1)

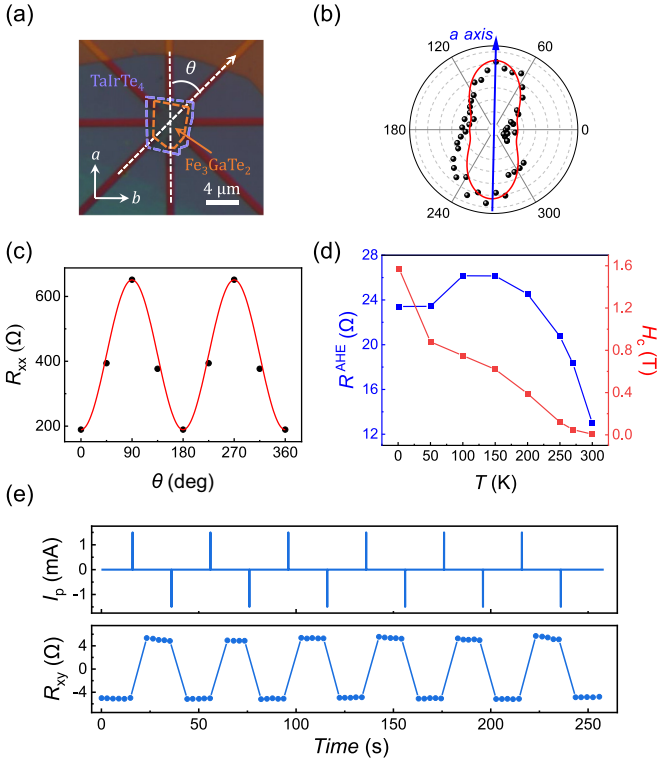

FIG. 11. (a) Optical image of device B. Fe$_3$GaTe$_2$ and TaIrTe$_4$ flakes are clearly labeled by orange and blue dashed borders, respectively. The angle $\theta$ is defined in the inset, where $\theta = 0°$ approximately corresponds to the $a$ axis and $\theta = 90°$ corresponds to the $b$ axis. (b) Polarized Raman spectroscopy, identifying the crystalline $a$ axis of TaIrTe$_4$. (c) Longitudinal resistance $R_{xx}$ as a function of $\theta$. (d) The anomalous Hall resistance (blue) and the coercive field (red) as a function of temperature. (e) Deterministic switching by a series of current pulses with the amplitude of $\pm 1.5$ mA at 300 K.

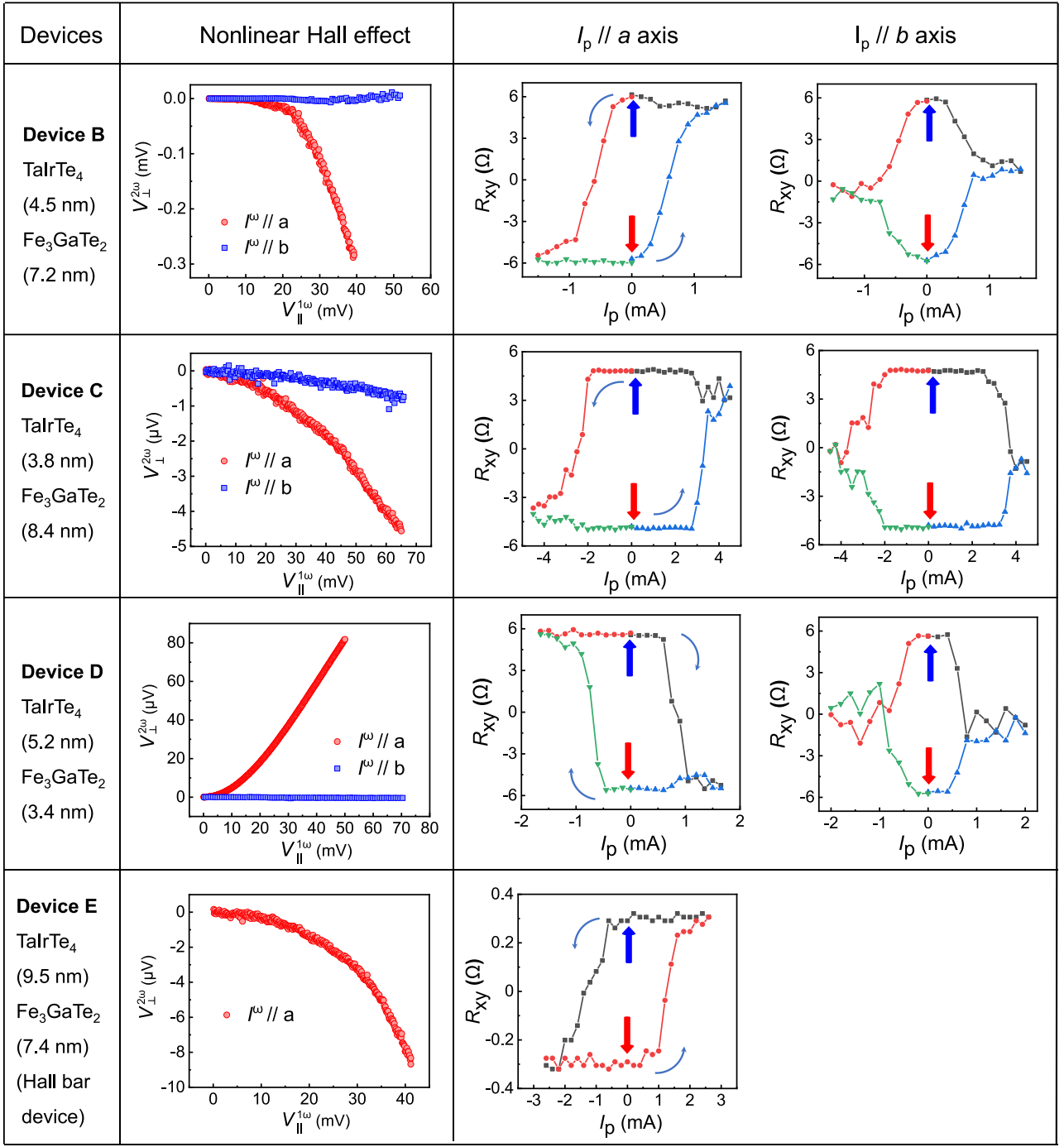

FIG. 12. Summary of the nonlinear Hall effect and field-free switching results in devices B–E.

where $\epsilon_c$ and $\epsilon_v$ are energies of the conduction band minimum and valence band maximum, respectively, $\mu_F$ is the Fermi energy, $g_e$ ($g_h$) is the density of states of electrons (holes), and $f_0$ is the Fermi-Dirac distribution. The corresponding carrier densities of electrons ($n_e$) and holes ($n_h$) of TaIrTe$_4$ are given in Fig. 10(c). Based on the carrier densities measured experimentally, the Fermi level of TaIrTe$_4$ is estimated to be near $\mu = 0.037$ eV [indicated by the blue dashed line in Fig. 10(c)] in our first-principles tight-binding Hamiltonian.

## APPENDIX C: ADDITIONAL DATA OF DEVICE B

Figure 11(a) shows the optical image of the TaIrTe$_4$/Fe$_3$GaTe$_2$ heterostructure for device B. The crystalline axes of TaIrTe$_4$ are aligned with the electrodes by identifying long, straight edges and they are confirmed by polarized Raman spectroscopy, as shown in Fig. 11(b). The longitudinal resistance $R_{xx}$ is plotted as a function of $\theta$ in Fig. 11(c). It is fitted by the equation $R_{xx}(\theta) = R_b \sin^2(\theta - \theta_0) + R_a \cos^2(\theta - \theta_0)$, where $R_a$, $R_b$ is the resistance along the $a$ axis and $b$ axis, respectively. Here $\theta_0$ corresponds to the angle misalignment between $\theta = 0°$ and the crystalline $a$ axis, and $\theta_0$ is $\sim 2°$ for this device. Figure 11(d) shows the temperature dependence of the coercive field and anomalous Hall resistance. The anomalous Hall resistance shows nonmonotonic dependence on temperature and the coercive field is weakened with increasing the temperature. Applying a train of pulse currents with an amplitude of $\pm 1.5$ mA can reliably switch the magnetizations of FGT, as seen in Fig. 11(e).

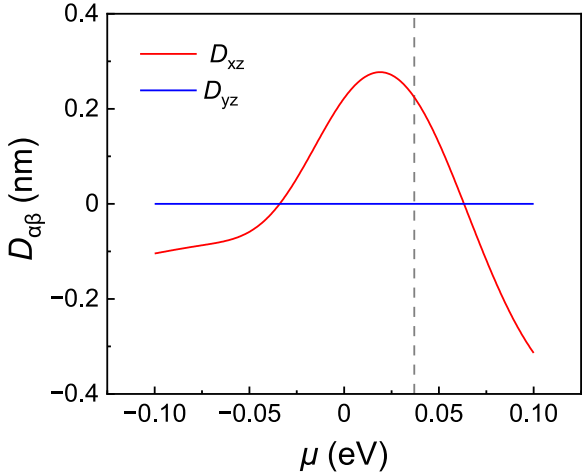

FIG. 13. Berry curvature dipole tensor $D_{yz}$ and $D_{xz}$ as a function of chemical potential $\mu$. The dashed line denotes Fermi energy of $\mu = 0.037$ eV. The temperature used for the calculations is set at 300 K. The coordinates $x$, $y$, and $z$ correspond to the $a$, $b$, and $c$ axes of TaIrTe$_4$, respectively.

## APPENDIX D: COMPARISON OF THE NONLINEAR HALL EFFECT AND FIELD-FREE SWITCHING RESULTS IN DEVICES B–E

See Fig. 12.

## APPENDIX E: THEORETICAL ANALYSES AND CALCULATIONS OF BERRY CURVATURE DIPOLE, ORBITAL, AND SPIN MAGNETOELECTRIC COEFFICIENTS

### 1. Berry curvature dipole

Berry curvature dipole, the dipole moment of the Berry curvature in momentum space, can be calculated as [43]

$$D_{ij} = -\int_{BZ} \frac{d^2k}{(2\pi)^2} \left[ \sum_n \partial_{k_i} \epsilon_{n\mathbf{k}} \Omega_{n\mathbf{k}}^j \partial_{\epsilon_{n\mathbf{k}}} f_{n\mathbf{k}}^{(0)} \right], \quad (E1)$$

where $i$ and $j$ represent $x$, $y$, or $z$, $\epsilon_{n\mathbf{k}}$ is the energy eigenvalue of the band $n$ at $\mathbf{k}$, $f_{n\mathbf{k}}^{(0)}$ is the Fermi distribution, and the Berry curvature can be found as

$$\mathbf{\Omega}_{n\mathbf{k}} = -\text{Im}\langle \nabla_\mathbf{k} u_{n\mathbf{k}} | \times | \nabla_\mathbf{k} u_{n\mathbf{k}} \rangle, \quad (E2)$$

where $u_{n\mathbf{k}}$ is the periodical part of the Bloch state of the band $n$ at $\mathbf{k}$. For a time-reversal symmetric system with broken inversion symmetry, the second-order nonlinear Hall effect can be originated from BCD. When applying an alternating electric field $\mathbf{E}^\omega$ and measuring the second-harmonic nonlinear response in the transverse direction, the induced second-harmonic Hall current is [43]

$$j_\alpha^{2\omega} = -\varepsilon_{\alpha\mu\gamma} \frac{e^3\tau}{2(1+i\omega\tau)\hbar^2} D_{\beta\mu} E_\beta^\omega E_\gamma^\omega, \quad (E3)$$

where $\varepsilon_{\alpha\mu\gamma}$ is the Levi-Civita symbol, $e$ is the electron charge, $\hbar$ is the reduced Planck constant, and charge $\tau$ is the relaxation time.

In our study, due to the presence of mirror symmetry $\mathcal{M}_a$ and broken glide mirror symmetry in thin-layer TaIrTe$_4$, the BCD tensor $D_{yz}$ is constrained to be zero while $D_{xz}$ is allowed, which indicates that an alternating electric field in the $x - y$ plane ($a - b$ plane) with a component along the $a$ axis of few-layer TaIrTe$_4$ should produce an in-plane nonlinear Hall voltage. With the first-principles calculations, we have studied the Berry curvature dipole in few-layer TaIrTe$_4$. Berry curvature dipole tensors as a function of chemical potential are shown in Fig. 13, from which we can extract $D_{xz}$ as 0.22 nm at the Fermi level (dashed line in Fig. 13).

### 2. Orbital and spin magnetoelectric coefficients

The electric field induced orbital and spin magnetization can be described as $\mathbf{M}_{\text{orb}} = (\frac{e^2\tau}{2\hbar^2})\alpha^{\text{orb}}\mathbf{E}$ and $\mathbf{M}_{\text{spin}} = (\frac{e\tau}{\hbar})\alpha^{\text{spin}}\mathbf{E}$, respectively, where $\mathbf{E}$ is the external electric field, $e$ is the electron charge, $\tau$ is the relaxation time, and $\alpha^{\text{orb}}$ ($\alpha^{\text{spin}}$) is the orbital (spin) magnetoelectric coefficient. The orbital (spin) magnetoelectric coefficient is evaluated by the following formula [23,50],

$$\alpha_{ij}^{\text{orb(spin)}} = \int_{BZ} \frac{d^2k}{(2\pi)^2} \sum_n m_{n\mathbf{k}}^{j,\text{orb(spin)}} \partial_{k_i} \epsilon_{n\mathbf{k}} \partial_{\epsilon_{n\mathbf{k}}} f_{n\mathbf{k}}^{(0)}, \quad (E4)$$

where $\mathbf{m}_{n\mathbf{k}}^{\text{orb(spin)}}$ is the orbital (spin) magnetic moment of each Bloch state. For the spin magnetic moment, $\mathbf{m}_{n\mathbf{k}}^{\text{spin}} = -\langle \partial_\mathbf{k} u_{n\mathbf{k}} | \frac{1}{2} g\mu_b \boldsymbol{\sigma} | \partial_\mathbf{k} u_{n\mathbf{k}} \rangle$, where $g \approx 2$ is the spin $g$ factor of the electron, $\boldsymbol{\sigma}$ is the Pauli operator, and $\mu_b$ is the Bohr magneton, while for the orbital magnetic moment, $\mathbf{m}_{n\mathbf{k}}^{\text{orb}}$ is described as $\mathbf{m}_{n\mathbf{k}}^{\text{orb}} = \text{Im}\langle \partial_\mathbf{k} u_{n\mathbf{k}} | \times (H_\mathbf{k} - \epsilon_{\mathbf{k}n}) | \partial_\mathbf{k} u_{n\mathbf{k}} \rangle$.

## APPENDIX F: DISCUSSION ON THERMAL EFFECT IN FIELD-FREE PM SWITCHING

When a large current is applied, it can generate Joule heating and increase the temperature of our device. Therefore, there is also a reduction in the coercive field of Fe$_3$GaTe$_2$, causing the shape of the anomalous Hall hysteresis loop to narrow at high current densities.

On the other hand, since Joule heating can cause a reduction in the coercive field, it will lower the energy barrier between the up and down magnetization of Fe$_3$GaTe$_2$, which could assist current-induced PM switching. However, the thermal effect is not the reason for achieving the deterministic PM switching. Firstly, although the coercive field of Fe$_3$GaTe$_2$ decreases when a large pulse current (critical current) is applied, the electron temperature increased by Joule heating remains below the Curie temperature of Fe$_3$GaTe$_2$ due to the observable hysteresis loop. Secondly, the resistance anisotropy in TaIrTe$_4$, where the resistance is minimum along the *a* axis and maximum along the *b* axis, has been observed. The Joule heat generated along the *a* axis is less compared to that generated along the *b* axis. However, there is no deterministic switching along the *b* axis, but it is clearly observed along the *a* axis. Thirdly, both current-induced PMA switching and AHE loop measurements exhibit a strong dependence on the crystal axis angle of TaIrTe$_4$, which is consistent with the mechanism of current-induced orbital magnetization.

Therefore, although the thermal effect may help to promote current-induced PM switching, it cannot play a deterministic role in the field-free magnetization switching.

---